\documentclass[12pt]{article}

\def\fnote#1#2{\begingroup\def\thefootnote{#1}\footnote{#2}\addtocounter{footnote}{-1}\endgroup}

\usepackage{amssymb}
\usepackage{makeidx}

\makeindex

\def\inbar{\vrule height1.5ex width.4pt depth0pt}
\def\IB{\relax{\rm I\kern-.18em B}}
\def\IC{\relax\,\hbox{$\inbar\kern-.3em{\rm C}$}}
\def\ID{\relax{\rm I\kern-.18em D}}
\def\IE{\relax{\rm I\kern-.18em E}}
\def\IF{\relax{\rm I\kern-.18em F}}
\def\IG{\relax\,\hbox{$\inbar\kern-.3em{\rm G}$}}
\def\IH{\relax{\rm I\kern-.18em H}}
\def\II{\relax{\rm I\kern-.18em I}}
\def\IK{\relax{\rm I\kern-.18em K}}
\def\IL{\relax{\rm I\kern-.18em L}}
\def\IM{\relax{\rm I\kern-.18em M}}
\def\IN{\relax{\rm I\kern-.18em N}}
\def\IO{\relax\,\hbox{$\inbar\kern-.3em{\rm O}$}}
\def\IP{\relax{\rm I\kern-.18em P}}
\def\IQ{\relax\,\hbox{$\inbar\kern-.3em{\rm Q}$}}
\def\IR{\relax{\rm I\kern-.18em R}}
\def\IT{\relax{\rm I\kern-.18em T}}
\def\ZZ{\relax{\sf Z\kern-.4em Z}}

\def\a{\alpha}   \def\b{\beta}    \def\g{\gamma}  
\def\e{\epsilon} \def\G{\Gamma}     
\def\L{\Lambda}    \def\Om{\Omega} \def\si{\sigma}

\def\cD{{\cal D}}  
 \def\cH{{\cal H}} \def\cI{{\cal I}}
  
 \def\cN{{\cal N}} \def\cO{{\cal O}}

   \def\tN{{\tilde N}}

\def\tPhi{\tilde \Phi}

  \def\wtN{{\widetilde N}}

  \def\rmM{{\rm M}}  \def\rmN{{\rm N}}

       \def\rmCHL{{\rm CHL}}

 \def\rmHet{{\rm Het}}     
 
       \def\rmIIA{{\rm IIA}}   \def\rmIm{{\rm Im}}

       \def\rmMS{{\rm MS}}

       \def\rmSk{{\rm Sk}}     \def\rmSL{{\rm SL}}
 \def\rmSO{{\rm SO}}      \def\rmSp{{\rm Sp}}

     \def\rmSym{{\rm Sym}}

\def\rmdd{{\rm dd}}            \def\rmdet{{\rm det}}
\def\rmdiag{{\rm diag}}   \def\rmdim{{\rm dim}}     \def\rmdom{{\rm dom}}

     \def\rmmod{{\rm mod}}

\def\rmrk{{\rm rk}}

 \def\rmtr{{\rm tr}}
     \def\rmwt{{\rm wt}}

\def\afrak{{\mathfrak a}} \def\cfrak{{\mathfrak c}}
 
 \def\mfrak{{\mathfrak m}}

 \def\mathC{{\mathbb C}} 
  \def\mathN{{\mathbb N}}

\def\mathQ{{\mathbb Q}} \def\mathR{{\mathbb R}}
\def\mathZ{{\mathbb Z}}

\def\fnote#1#2{\begingroup\def\thefootnote{#1}\footnote{#2}\addtocounter{footnote}{-1}\endgroup}

\def\beq{\begin{equation}}
\def\eeq{\end{equation}}
\def\bea{\begin{eqnarray}}
\def\eea{\end{eqnarray}}
\def\llea#1{\label{#1}\eea}
\def\lleq#1{\label{#1}\eeq}
\let\nn=\nonumber
\def\tabroom{\hbox to0pt{\phantom{\Huge A}\hss}}
\def\notin{\ \hbox{{$\in$}\kern-.51em\hbox{/}}}

\def\ra{{\rightarrow}}
\def\lra{\longrightarrow}

 \def\vphi{\varphi}

  \def\E1Fq{E_1/\IF_q}

\def\notdiv{{\relax{~|\kern-.34em /~}}}

\def\boxit#1{\vbox{\hrule height1pt\hbox{\vrule width1pt\kern0.3cm
\vbox{\kern0.3cm\hbox{$\displaystyle#1$}\kern0.3cm}\kern0.3cm\vrule
width1pt}\hrule height1pt}}

\begin{document}

\hfill {\bf MPP-2011-121}

%\hfill \today

\baselineskip=20pt
\parindent=0pt

\vskip .9truein

 \centerline{\large {\bf Automorphic Black Holes as Probes of Extra Dimensions}}

\vskip .3truein

\centerline{\sc Kayleigh Cassella$^1$\fnote{*}{Present address: Dept. of Physics, University of California, Berkeley;
                                                                cassella@berkeley.edu}
               and Rolf Schimmrigk$^2$\fnote{$\diamond$}{On leave from Indiana University South Bend; netahu@yahoo.com,
               rschimmr@iusb.edu}}

\vskip .2truein

\centerline{ $^1$Department of Physics}

\centerline{ Indiana University South Bend}

\centerline{ 1700 Mishawaka Ave., South Bend, IN 46634, USA}

\vskip .2truein

 \centerline{ $^2$Max-Planck-Institut f\"ur Physik}

\centerline{ F\"ohringer Ring 6, 80805 M\"unchen, Germany}

 \vskip .5truein

\baselineskip=17pt

 \centerline{\bf Abstract:}
 \begin{quote}
 Recent progress in the understanding of the statistical nature of black hole
 entropy shows that the counting functions in certain classes of
 models are determined by automorphic forms of higher rank. In this
 paper we combine these results with Langlands' reciprocity
 conjecture to view black holes as probes of the geometry
 of spacetime. This point of view can be applied in any framework leading to
 automorphic forms, independently of the degree of supersymmetry of the models.
 In the present work we focus on the class of Chaudhuri-Hockney-Lykken compactifications
  defined as quotients associated to $\mathZ_N$ groups. We show
  that the black hole entropy of  these CHL$_N$ models can be derived from elliptic motives, 
  thereby providing
 the simplest possible geometric building blocks of the Siegel type entropy count.
 \end{quote}

\renewcommand\thepage{}
\newpage
\parindent=0pt

\baselineskip=16pt
\parskip=.1truein

\renewcommand\thepage{}
\newpage
\parindent=0pt

 \pagenumbering{arabic}

 \vfill \eject

 \baselineskip=19pt
 \parskip=4pt

 \tableofcontents

\vfill \eject

\baselineskip=21.1pt
\parskip=.15truein

\section{Introduction}

Black holes can be viewed as objects that encode structural
information about the theories in which they are embedded. As such
they provide probes that can be used to ask what exactly we could
learn about the ambient physical theory if we were able to perform experiments with them.
One particular focus in black
hole physics over the past four decades has been the problem of a
fundamental understanding of their entropy, in particular its
statistical interpretation. The purpose of this paper is to
 address the question what kind of information is encoded in the
 automorphic entropy functions that have recently been constructed for certain
 types of black holes. The idea developed here is to view black
 hole entropy as a probe that is sensitive to the geometric structure of
 the extra dimensions predicted by string theory.

The microscopic understanding of black hole entropy has made great
progress in the past few years. In the context of $\cN=4$
compactifications these results have lead to partition functions
that provide a Boltzmann count of the numbers of states. A class
of models that has received much attention is based on
Chaudhuri-Hockney-Lykken constructions obtained via quotients with
respect to cyclic groups $\mathZ_N$, denoted here by CHL$_N$
\cite{chl95}. Dijkgraaf-Verlinde-Verlinde \cite{dvv96}, Jatkar-Sen
\cite{js05}, and Govindarajan-Krisnhna \cite{gk09} have shown that
the entropy of certain types of dyonic black holes is completely
determined by Siegel modular forms $\Phi^N$ of genus two, the
structure of which depends on the quotient group
$\mathZ_N$ of the CHL$_1$ compactification manifold $T^6$ (see also ref.
 \cite{ssy05}). Siegel
modular forms define a special class of automorpic forms that
generalize to the symplectic groups $\rmSp(2n,\mathZ)$ the
classical modular forms derived from congruence subgroups of the
modular group $\rmSp(2,\mathZ)\cong \rmSL(2,\mathZ)$. It has been known
for more than a
century that certain modular forms have a geometric origin, and generalizations
to automorphic forms
 have been discussed more recently. Such experimental results have led to a web of
conjectures by Taniyama, Shimura, Weil, and later Langlands and
others, that raise the hope that at least certain classes of
automorphic forms are of geometric origin. In string theory the
notion of geometric automorphic forms has previously been used to
provide a construction of the compactified geometry directly in
terms of modular forms on the worldsheet, leading to a framework
that realizes the idea of an emergent spacetime in string theory
\cite{rs08}.

The geometric construction of automorphic forms and their
associated representations provides meaning to the general
question whether it is possible to deduce an underlying
irreducible geometric structure that leads to the automorphic
forms that appear in black hole entropy counting problems in
string theory, and if so, whether these geometric structures are
unique.

In this paper we address this problem in the context of the
CHL$_N$ theories and their associated Siegel modular forms.
 Not much has been proven about the geometric interpretation of Siegel
modular forms even in the special case of genus $n=2$, but we will
see that the conjectural framework of Langlands applied to Siegel
forms implies that it is not possible to find geometric structures
in the CHL$_N$ models that directly support the black hole Siegel
forms in the form usually envisioned. For this reason it is useful
to first analyze the precise structure of the CHL$_N$ type Siegel
forms $\Phi^N$ in more detail before addressing the question of
the geometric origin of these objects, and thus of the entropy of
the CHL$_N$ black holes. The first simplification that arises in
the context of the CHL$_N$ models is that the Siegel forms
$\Phi^N$ encoding the entropy of the CHL$_N$ type black holes are
lifts of simpler types of modular forms. The lifts relevant for
the CHL$_N$ black holes were first considered by Maa\ss~and
Skoruppa for modular forms of level one, and later generalized to
higher level. It was shown that the Siegel forms $\Phi^N$ of the
CHL$_N$ model belong
 to the so-called Maa\ss~Spezialschar, and are determined by classical
modular cusp forms $f^N \in S_w(\G_0(N))$ of level $N$ with
respect to the Hecke group $\G_0(N)$ of some weight $w$. These
classical forms are Hecke eigenforms that are determined by the
electric (or magnetic) BPS states, and lead to $\Phi^N$ by
composing two maps, the Skoruppa lift from classical forms to
Jacobi forms, and the Maa\ss~lift from Jacobi forms to Siegel
modular forms. These lifts are completely canonical, independent
 of $N$, hence the forms $f^N$ provide the key building blocks of
the dyonic black hole count. In the following the classical
modular forms $f^N$ will be called the Maa\ss-Skoruppa roots, or
the black hole roots, of the Siegel modular forms $\Phi^N$.

The problem of a geometric understanding of the black hole entropy
in CHL$_N$ models therefore translates into the problem of
understanding the geometric origin of the Maa\ss-Skoruppa roots.
It is this question that we address in this paper. We show that
while the motives naively associated to the Maa\ss-Skoruppa roots
are not physical, it is possible to reduce these modular forms
further, and to construct all the forms $f^N$ of the class of
CHL$_N$ models in terms of classical modular forms $f_2^\tN$ of
weight two, where the level $\tN$ is determined by $N$. Our lift
construction therefore implies that motivically these forms are
supported by elliptic curves $E_\tN$ of conductor $\tN$, which are
determined up to isogeny. The view of black holes as probes of the
geometry of spacetime raises the question in what detail the
entropy probes the extra dimensions. The fact that the dyon counts
of the CHL$_N$ models are determined by elliptic curves $E_\tN$
shows that the underlying geometry of the entropy is that given by
a single motive, not by the composite motivic structure expected
from more complicated compactification manifolds. In general
higher dimensional manifold are determined by several modular, or
automorphic, motives. Hence the fact that for each model there is
a single modular form shows that the CHL$_N$ black hole entropy
functions considered so far do not probe the full structure of the
compact geometry.

The outline of the paper is as follows. In Section 2 we briefly
describe the microscopic structure of CHL$_N$ black holes, in
particular the lift structure of the Siegel modular forms that is
relevant for this paper. In Section 3 we describe the general
motivic framework associated to Siegel modular forms, and in
Section 4 we derive the underlying motivic building blocks of the
Siegel forms that arise in the CHL$_N$ models. In Section 5 we
summarize our work, and in an Appendix we analyze the symmetry
structure of the forms that appear in our discussion.

\vskip .3truein

\section{Black hole entropy of the CHL$_N$ models}

\subsection{CHL$_N$ models}

The first step toward a generalization of $\cN=4$ black hole
entropy was taken by Jatkar and Sen \cite{js05}. These authors
formulated a proposal for the dyonic partition functions in the
class of Chaudhuri-Hockney-Lykken models CHL$_N$ for $N$ prime.
The compact manifolds in these models are quotient spaces with
respect to some cyclic group $\mathZ_N := \mathZ/N\mathZ$ of order
$N$
 \beq
 \rmCHL_N:~~~  \rmHet(T^6/\mathZ_N)  ~\cong ~ \rmIIA((K3\times
   T^2)/\mathZ_N).
 \eeq
 In the IIA frame the group $\mathZ_N$ acts via a symplectic automorphism
on the K3 factor and as an order $N$ shift on one of the 1-cycles
of the torus $T^2$.

The reason why Siegel modular forms arise in the CHL$_N$ models
can be traced to the duality group, which for the low energy
supergravity theory is given by
 \beq
 U_N(\mathR) ~=~ \rmSL(2,\mathR) \times \rmSO(6,r_N-6,\mathR),
 \eeq
 where $r_N$ is the rank of the gauge group of the CHL$_N$ theory. In the full string theory
 of the CHL$_N$ models this group is broken to the subgroup
 \cite{as05a}
 \beq
 U_N(\mathZ) ~=~  \G_1(N) \times \rmSO(6,r_N-6,\mathZ),
 \eeq
 where
 \beq
 \G_1(N) ~=~  \left\{\g \in \rmSL(2,\mathZ) ~{\Big |}~
         \g \equiv \left(\matrix{1 &*\cr 0 &1\cr}
                \right) (\rmmod~N)\right\}.
 \eeq
 Invariance under the duality group implies that the
 dyon degeneracies, a priori functions of the charges
 \beq
 \rmCHL_N:~~~~(Q_e,Q_m) ~\in ~ \L_N \oplus \L_N,
 \eeq
 where $\L_N=\L^{6,r_N-6}$ is a Narain lattice, depend only on
 the duality invariant norms, given by
 \beq
 (Q_e^2,~Q_m^2,~Q_eQ_m) ~\in ~
   \frac{2}{N}\mathZ \times 2\mathZ \times \mathZ.
 \lleq{charge-quantization}

 Physical quantities invariant under T-duality then should depend
 on the charges through their three invariant norms.

\subsection{Siegel automorphic forms of genus 2}

 The fact that there are three T-invariant norms suggests to introduce
  three chemical potentials, associated to $Q_m^2,Q_e^2, Q_eQ_m$, denoted here by
 $(\tau,\si,\rho)$. The partition function of the CHL$_N$ models are then expected
 to be expressed in terms of a 3-variable automorphic forms
 $\Phi(\tau, \si, \rho)$ as
 \beq
 Z(\tau, \si,\rho) ~=~ \frac{1}{\Phi(\tau,\si,\rho)} ~=~
 \sum_{k,\ell,m} d(k,\ell,m) q^kr^\ell s^m,
 \lleq{siegel-partition-function}
 where  $q=e^{2\pi i \tau}, r=e^{2\pi i \si}, s=e^{2\pi i \rho}$,
 and the Fourier expansion of the form can be written as
 \beq
 \Phi(q,r,s) ~=~ \sum_{k,\ell,m} g(k,\ell,m) q^kr^\ell s^m,
 \lleq{genus-2-expansion}
 with $k,\ell,m$ are integers determined by the T-duality
 invariant norms in eq. (\ref{charge-quantization}).
 The detailed structure of the form $\Phi$ will depend on the
 structure of the models considered.

 A well-known class of automorphic forms are Siegel modular forms of genus $n$,
 defined as functions
  \beq
     \Phi_{w}: ~\cH_n ~\lra \mathC
 \eeq
 on the Siegel upper halfplane
  \beq
 \cH_n ~=~ \left\{T \in M_n(\mathC) ~{\Big |}~
   T~{\rm symmetric, with~positive-definite~imaginary ~part} \right\}
 \eeq
 of dimension $\rmdim_\mathC \cH_n = \frac{n}{2}(n+1)$.
 Siegel modular forms of genus 2 are therefore defined on a
 three-dimensional space,
 and it is natural to check whether such Siegel forms
 provide a useful framework for CHL$_N$ models.

 Like classical modular forms, Siegel modular forms are characterized
 by a weight $w$ and a level $N$, determined
 by the relevant congruence subgroup $\G^{(n)}(N)$ of the symplectic
 group $\rmSp(2n,\mathZ)$.
 The functions $\Phi_w$ satisfy a scaling behavior with respect to
 elements $M\in \G^{(n)}(N) \subset \rmSp(2n,\mathR)$, where
 the action of $M$ on $\cH_n$ is defined as
 \beq
 MT ~=~ \left(\matrix{A &B\cr C &D\cr}\right) T ~=~ (AT+B)(CT+D)^{-1}.
 \eeq
 In the following analysis only scalar Siegel modular
 forms are needed, for which the transformation behavior is given
 by
 \beq
 \Phi_w(MT) ~=~ j_w(M,T) \Phi_w(T) ~=~ \rmdet(CT+D)^w \Phi_w(T),
 \eeq
 where $w$ is assumed to be integral.

 The full symplectic group $\rmSp(2n,\mathZ)$ is too large to allow
 many Siegel forms at fixed
 weight, and it is necessary to consider congruence subgroups of the
 full symplectic group.
 This is similar to the case of classical modular forms, where the full
 modular group
 $\rmSL(2,\mathZ)$ is too restrictive as well to allow for many
 interesting modular forms
 at fixed weight. As in the classical case there are different types
 of congruence groups
  that are of interest for different questions. In the context of
  CHL$_N$ models the groups
 of interest are $\G_0^{(n)}(N) \subset \rmSp(2n,\mathZ)$, defined for
 arbitrary genus $n$ as
  \beq
 \G_0^{(n)}(N) ~=~ \left\{\left(\matrix{A &B\cr C &D\cr}\right) \in \rmSp(2n,\mathZ)
         ~{\Big |}~ C\equiv 0(\rmmod~N)\right\}.
 \eeq
 These groups generalize to $\rmSp(4,\mathZ)$ the Hecke congruence subgroup
 $\G_0(N) \subset \rmSL(2,\mathZ)$ of the full modular group $\rmSL(2,\mathZ)$.

 The Fourier expansion of $\Phi_w(T)$ takes the form
 \beq
 \Phi_w(T) ~=~ \sum_{\stackrel{0\leq U^t=U}{\rm semi-integral}}
 g(U) e^{2\pi i \rmtr (UT)},
 \eeq
 where semi-integral means that the diagonal entries of $U$ are
 integers, while the off-diagonal entries are either integers of
 half-integers. For $n=2$ the variables of the Siegel upper plane take the form
 \beq
 T ~=~ \left(\matrix{\tau &\rho \cr \rho &\si}\right) \in \cH_2
 \eeq
 with $\tau, \si \in \cH_1$, i.e.
 \beq
 \rmIm(\tau) ~>~0,
 ~~~~\rmIm(\si)~>~0,
 ~~~~ \rmIm(\tau)\rmIm(\si) ~>~ \rmIm(\rho)^2.
 \eeq
  The functional dependence is often written as
  $\Phi_w(\tau, \si, \rho) ~=~ \Phi_w(T)$, and the Fourier expansion
 can be expressed via
 $$
 U ~=~ \left(\matrix{k &m/2 \cr  m/2  &\ell \cr}\right),
 ~~~{\rm with} ~~~k,\ell,m \in \mathZ, ~k,\ell\geq 1,
                            m^2 < 4k\ell,
 $$
 as
 \beq
  \Phi_w(T)
  ~=~ \sum_{\stackrel{k,\ell\in \mathN,m\in \mathZ}{k,\ell, 4k\ell-m^2 > 0}}
               g(k,\ell,m) q^k r^\ell s^m.
 \eeq
 The Fourier coefficients $g(k,\ell,m)$ determine the degeneracies $d(k,\ell,m)$
 via the partition
 function (\ref{siegel-partition-function}).

\subsection{The Maa\ss-Skoruppa lift for CHL$_N$ Siegel modular forms}

We have seen that the duality invariance of the CHL$_N$ model suggests
to look for three-dimensional automorphic forms, leading to Siegel
modular forms of genus two as the simplest candidates. As noted
above, Siegel forms are characterized like classical modular forms
by their weight $w^N=w(\Phi^N)$ and their level. It turns out that
their weight is given in terms of the rank $r_N$ of the gauge
group of the CHL$_N$ model as
 \beq
 w^N ~=~ \frac{1}{2}(r_N-8).
 \eeq
 We will identify the Siegel forms $\Phi^N$ by their level instead
 of the weight. The rank $r_N$ of the $\mathZ_N-$model depends not
 only on the order
 of the quotient group $\mathZ_N$ but also on the precise form of the action. For
the models considered here they are given by in Table 1.
 %\cite{a95}
 \begin{center}
 \begin{tabular}{l| c c c c c c c c}

 $N$   &1   &2  &3  &4   &5  &6  &7  &8 \tabroom \\
 \hline

 $r_N$ &28  &20 &16 &14  &12 &12 &10  &10  \tabroom \\

 \hline
 \end{tabular}
 \end{center}
 \centerline{{\bf Table 1.}~{\it Ranks $r_N$ of the CHL$_N$
 models.}}

The embedding of the chemical potentials into the genus two Siegel
upper halfplane $\cH_2$ shows that in the limit $\rho \ra 0$ the
Siegel form $\Phi^N(\tau,\si, \rho)$ should factorize as
 \beq
 \Phi^N(\tau,\si,\rho) ~~\stackrel{\rho\ra 0}{\lra} ~
   \sim \a(\rho) f^N(\tau) g^N(\si),
 \lleq{siegel-diagonal}
 where $f^N(\tau)$ corresponds to purely electrically charged
 states, while $g^N(\si)$ corresponds to purely magnetically
 charged states. Electro-magnetic duality leads to $f^N ~=~ g^N$.

 The factorization (\ref{siegel-diagonal}) along the diagonal
 suggests that the Siegel modular forms
 describing the CHL$_N$ models can be constructed as lifts of
 classical modular forms $f^N$. Such lifts have been constructed by
 Maa\ss~\cite{m79} and Skoruppa \cite{nps92} for the full modular group,
 and extensions for congruence groups have been discussed in refs.
 \cite{mrv93, cg08}.
 We will call this construction the Maa\ss-Skoruppa lift, or additive lift.
 It is obtained via a two-step construction, the Skoruppa lift $\rmSk(f^N)$
 from classical modular forms $f^N$ to Jacobi forms $\vphi^N$, and the Maa\ss~lift
 $\rmM(\vphi^N)$ from Jacobi forms $\vphi^N$ to Siegel forms $\Phi^N$
 \beq
 f^N ~\stackrel{\rmSk}{\lra} ~ \vphi^N ~ \stackrel{\rmM}{\lra} ~ \Phi^N = \rmMS(f^N).
 \lleq{ms-lift}
 We briefly outline these two lifts because
 this induction of Siegel forms by their Maa\ss-Skoruppa roots
 will be the starting point for our motivic interpretation
 described in the next section.

\subsubsection{The Skoruppa lift}

The first step of the additive lift from classical modular forms
to CHL$_N$ black hole Siegel forms is based on a result first
shown by Skoruppa \cite{nps92} for level one forms, and later
extended by Cl\'ery and Gritsenko \cite{cg08} to modular forms of
higher level. This construction implements a map from classical
cusp forms to Jacobi forms.

Jacobi modular forms of weight $w$ and index $\ell$ are maps
 \beq
 \vphi_{w,\ell}: ~~\cH \times \mathC ~\lra ~ \mathC
 \eeq
 such that for any element $\g = \left(\matrix{a &b\cr c &d\cr}\right)$ in some
  congruence subgroup $\g \in \G \subset \rmSL(2,\mathZ)$
 \beq
  \vphi_{w,\ell}\left(\frac{a\tau +b}{c\tau +d}, \frac{\rho}{c\tau + d}\right)
  ~ = ~ (c\tau+d)^w e^{2\pi i \ell c \frac{\rho^2}{(c\tau + d)}} \vphi_{w,\ell}(\tau,\rho).
 \eeq
 %there is a typo on p1 of Eichler-Zagier85
  Furthermore there is a transformation of the group
 $\mathZ^2$, acting and transforming like
 \beq
 \vphi_{w,\ell}(\tau, \rho + \a \tau + \b)
    ~=~ e^{-2\pi i \ell \a(\a\tau  + 2\rho)} \vphi_{w,\ell}(\tau,\rho),~~~~~
         (\a,\b)\in \mathZ.
 \eeq
 Jacobi cusp forms admit a Fourier expansion as
 \beq
 \vphi_{w,\ell}(\tau,\si)
  ~=~ \sum_{k\in \mathZ_{\geq 0}}
         \sum_{\stackrel{m\in \mathZ}{4k\ell - m^2 > 0}} c(k,m) q^k
         s^m
 \lleq{jacobi-fourier-expansion}
 while for general Jacobi forms the expansion is restricted by $4k\ell-m^2\geq 0$.
 The space of Jacobi forms of weight $w$ and index $\ell$ with
 respect to some congruence group $\G \subset \rmSL(2,\mathZ)$
 will be denoted by $J_{w,m}(\G)$, or simply $J_{w,m}$.
 \index{$J_{w,m}(\G)$}

The map sending cusp forms $f\in S_w(\G_0(N),\e)$ of weight $w$,
level $N$, and character $\e$ to Jacobi forms
 \beq
 \rmSk:~ S_{w+2}(\G_0(N),\e) ~\lra ~ J_{w,1}
 \eeq
 will be called the Skoruppa map. It is defined by multiplication with the prime form
 \beq
 K(\tau,\rho) ~=~ \frac{\vartheta_1(\tau,\rho)}{\eta^3(\tau)},
 \eeq
 given in terms of the Dedekind eta function
 \beq
 \eta(q) ~=~ \prod_{n\geq 1} (1-q^n),
 \eeq
and the theta series $\vartheta_1(\tau,\si)$ defined as
 \beq
 \vartheta_1(q,s) ~=~ \sum_{n\in \mathZ} (-1)^n
       q^{\frac{1}{8}(2n+1)^2} s^{n+\frac{1}{2}}.
 \eeq
 The lift is then given by
 \beq
    \vphi_{w,1}(\tau,\rho) := K^2(\tau,\rho) f(\tau).
 \eeq
 The square of the prime form is one of the generators of the
 space of weak Jacobi forms of even weight and integral index
 $K^2(\tau,\rho) ~=~ \vphi_{-2,1}(\tau,\rho)$.

\subsubsection{The Maa\ss-Skoruppa lift}

The second step of the additive lift construction uses a result
shown first by Maa\ss~for level one modular forms, and later
extended by Manickam, Ramakrishnan and Vesudan \cite{mrv93} to
higher level.
 This Maa\ss~lift constructs the Fourier coefficients $g(k,\ell,m)$
 of the expansion (\ref{genus-2-expansion})
 of the Siegel modular forms $\Phi_w(q,r,s)$
 in terms of the coefficients $c(k,m)$ of the Fourier expansion
 (\ref{jacobi-fourier-expansion}) of a Jacobi form of weight $w$ and index 1.
 For a Jacobi form of weight $w$ and index $\ell$ the coefficients $c(k,m)$ depend
 on $k,m$ only via the combination $(4k\ell-m^2)$. The Fourier coefficients
 of Siegel forms in the Maa\ss~subspace are then given as
 \beq
 g(k,\ell,m) ~=~ \sum_{d|(k,\ell,m)} \chi(d) d^{w-1}
 c\left(\frac{k\ell}{d^2},\frac{m}{d}\right),
 \eeq
 where $\chi$ is a character that is either trivial or given by a
 Dirichlet character.

A more conceptual formulation of the Maa\ss~lift can be obtained
by noting that the Hecke operators $T_m$ acting on Jacobi forms
$\vphi_{w,\ell}(q,s)$ of weight $w$ an index $\ell$ produce Jacobi
forms $\vphi_{w,\ell+m}$ of the same weight and index $\ell+m$
 \beq
 T_m \vphi_{w,\ell} ~=~ \vphi_{w,\ell+m}.
 \eeq
 With these operators one can then generate the
 Fourier-Jacobi expansion of the Siegel modular form as
 \beq
 \Phi_w(q,r,s)
   ~=~ \sum_\ell \vphi_{w,\ell}(q,s) r^\ell
   ~=~ \sum_\ell (T_\ell \vphi_{w,1}) (q,s) r^\ell.
 \eeq

 The Maa\ss-Skoruppa lift, obtained by combining the
 Maa\ss~lift with the Skoruppa lift $\rmM \circ \rmSk$, therefore
 maps classical modular forms to Siegel forms of genus 2
 \beq
 \Phi_w ~=~ \rmM(\vphi_{w,1}) ~=~ \rmM(\rmSk(f_{w+2})) ~=:~ \rmMS(f_{w+2}).
 \eeq
 In the context of the CHL$_N$ models we will characterize the
 modular forms by their model index $N$ rather than the weight.
 The weight $(w+2)$ of the forms $f_{w+2}=f^N$ is determined by the
 order $N$ of the CHL$_N$ group, as described in the next
 subsection.

\subsection{A simple choice for the electric (magnetic) modular forms}

 The present subsection describes a rationale that identifies unique
 candidates for the CHL$_N$ model Maa\ss-Skoruppa roots in a very simple way.
 First, recall that the S-duality of electromagnetism generalizes
 to the toroidal model as $\rmSL(2,\mathZ)$. It turns out that in
 the $N=1$ model considered in \cite{dvv96} the electric modular form $f^1(q)$ is
 in fact the unique eta product with the respect to the full
 modular group.

Recall next that the duality group $\rmSL(2,\mathZ)$ of the $N=1$
model is broken for higher $N$ to the congruence group $\G_1(N)
\subset \rmSL(2,\mathZ)$ \cite{as05a}.
 Assuming that the generalization of the electric BPS counting
 form $f^N(q)$ for arbitrary $N>1$ generalizes in the simplest possible way the
 $N=1$ partition function  $f^1(q) = \eta(q)^{24}$  leads to a simple guess:
 it is natural to expect that the generalization $f^N(q)$ of $f^1(q)$ is given
 by modular forms of level $N$. For each prime order $N=p \in \{2,3,5,7\}$ there is a
unique candidate cusp form $\eta-$product with the appropriate
level and integral weight. These forms are given can be written in
closed form in terms of the Dedekind eta function as
 \beq
 f^N(q) ~=~ \eta(q)^{w+2}\eta(q^N)^{w+2} ~\in ~ S_{w+2}(\G_0(N), \e_N),
 \lleq{js-list}
 where for $N=1,2,3,5,7,11$ the resulting weight is given by
 $w+2 = \frac{24}{(N+1)}$, and $\G_0(N) \subset \rmSL(2,\mathZ)$ is Hecke's congruence group.
 The character $\e_N$ is only nontrivial
 for level $N=7$, in which case it is given by the Legendre
 character $\e_7(d) = \chi_{-7}(d) = \left(\frac{-7}{d}\right).$
 In general, the Legendre character is defined as
 \beq
 \chi_N(p) ~=~ \left(\frac{N}{p}\right)
  ~=~ \left\{ \begin{tabular}{r l}
             1  &if $x^2\equiv N(\rmmod~p)$ is solvable \\
             $-1$ &if $x^2\equiv N(\rmmod~p)$ is not solvable \tabroom \\
             0  &if $p|N$. \tabroom \\
             \end{tabular}
     \right\}
 \eeq

 For the composite values $N=4,6,8$ that complete the CHL$_N$
 sequence of models the quotient $24/(N+1)$ is neither integral nor half-integral.
 It is natural to extend the above sequence for prime order  by considering forms
 of weight
 \beq
 w+2 := \left\lceil \frac{24}{N+1}\right\rceil,
 \lleq{weight-of-ms-roots}
 where $\lceil a \rceil$ denotes the next largest integral number obtained from the
 rational number $a$.
 For  $N=4,6,8$ these forms therefore are of weight $5,4,3$, respectively.
 Extending
 furthermore the expectation that the order of the group again determines
 the level of the
 modular form leads to unique candidates of eta products given by
 \bea
  f^{4}(\tau) &=& \eta(\tau)^4 \eta(2\tau)^2 \eta(4\tau)^4 ~\in ~
                 S_5(\G_0(4),\chi_{-1})\nn \\
  f^{6}(\tau) &=& (\eta(\tau)\eta(2\tau)\eta(3\tau)\eta(6\tau))^2 ~\in ~ S_4(\G_0(6)) \nn \\
  f^{8}(\tau) &=& \eta(\tau)^2\eta(2\tau)\eta(4\tau)\eta(8\tau)^2 ~\in~ S_3(\G_0(8),\chi_{-2}).
 \llea{gk-list}
 The characters are again given by Legendre characters.
 The forms obtained in eqs. (\ref{js-list}) and (\ref{gk-list})
  are precisely the forms proposed by Jatkar and Sen \cite{js05} for prime
  orders, and
 by Govindarajan and Krishna \cite{gk09} for the composite orders.

Using as input for the Skoruppa lift the forms
$f^{N}\in S_{w+2}(\G_0(N),\e_N)$, with weights $(w+2)$ given by
 (\ref{weight-of-ms-roots}), leads to Jacobi forms $\vphi^N(q,s)$
of weight $w$ and index 1. The Maa\ss~lift of $\vphi^N(q,s)$ then
 leads to Siegel modular forms $\Phi^N(q,r,s) \in S_w(\G_0^{(2)}(N))$.

The final step of the Siegel formulation of the microscopic
entropy is motivated by the map between the diagonal and dominant
zero divisors of the Siegel form. This map is obtained by an
$\rmSp(4,\mathZ)$ matrix $M_\rmdd = \left(\matrix{A_\rmdd
&B_\rmdd\cr C_\rmdd &D_\rmdd\cr}\right)$ chosen such that the
image of the diagonal divisor $\cD_\rmdiag :=\{\rho^2=0\}$ is
essentially the dominant divisor given by $\cD_\rmdom :=\{\rho^2 -
\rho - \tau \si=0\}$.
 The transformation behavior of the weight $w$ Siegel form $\Phi_w$ then
 suggests the introduction of the form
 \beq
    \tPhi_w(T) := \rmdet(C_\rmdd T'+D_\rmdd)^w \Phi_w(T'),
 \lleq{siegel-tilde}
 where the coordinates $T'$ are defined as $T = M_\rmdd T'$.
 With this form the degeneracies are defined by
 \beq
 d(Q_e,Q_m) ~=~ (-1)^{Q_eQ_m+1} \int dT \frac{e^{-\pi i Q^t TQ}}{\tPhi_w(T)},
 \lleq{siegel-degeneracies}
 where $dT=d\tau d\si d\rho$, $Q = \left(\matrix{Q_m\cr Q_e\cr}\right)$, and
 $Q_e,Q_m$ are the charges
 associated to the gauge fields of the CHL$_N$ gauge fields. The
 microscopic entropy of black holes of charge $(Q_e,Q_m)$ is then
 defined by
 \beq
 S^{\rm mic}(Q_e,Q_m) ~=~ \ln ~|d(Q_e,Q_m)|.
 \lleq{siegel-entropy}
This entropy has been shown to agree in certain approximations with the
macroscopic entropy
derived e.g. via the OSV framework, along the lines of ref. \cite{cdwkm04}.

\vskip .3truein

\section{Black holes as probes of spacetime geometry}

The idea of using black holes as probes of spacetime geometry
motivates the question what precisely the information is that is
encoded in their entropy. Specifically, we can ask whether we can
learn something about the structure of the extra dimensions from
the entropy of black holes. In this section we formulate the
framework in which these vague questions can be made precise.

The microscopic entropy framework provided by the Siegel forms via
(\ref{siegel-tilde}), \ref{siegel-degeneracies}),
(\ref{siegel-entropy}) shows that the information encoded in the
entropy is completely determined by the Siegel form $\Phi^N$. Once
this form is has been identified, the degeneracies and the entropy
are known. To determine the geometric origin of the black hole
entropy thus means to determine the geometric origin of the Siegel
modular forms.

In the context of the CHL$_N$ models, viewed in the type II string
framework as compactifications $\rmCHL_N = \rmIIA(X_N)$ on
varieties of the type
 $$
 X_N ~=~ (K3\times E)/\mathZ_N,
 $$
 it is natural to ask whether the modular forms that arise in the
 black hole entropy functions can be used as probes that would allow to deduce
 information about the compact geometry if one could perform experiments
 with CHL$_N$ black holes in the laboratory.
 With a physical probe that is highly sensitive to the details of
 the ambient spacetime one might hope to be able to reconstruct
completely the precise structure of this geometry. We will
 show that for the entropy of CHL$_N$ black holes this is not the
case. The relevant geometric information that is encoded in the
entropic Siegel modular forms is motivic (see below) in the sense
that a single geometric structure suffices to determine these
forms completely. In general a variety, in particular varieties of
the form $X_N$ describing the compactification of the CHL$_N$
models, lead to several motivic building blocks that characterize
the manifold completely, while a single motive can often be
embedded into different varieties.

\subsection{Automorphic motives as geometric building blocks}

The problem of finding the geometric origin of automorphic forms
leads to the notion of motives, certain substructures of manifolds
that are reflected in the cohomology of a variety. Manifolds are
not to be viewed as single monolithic objects but instead as a
coherent structure of several different building blocks, where the
same building blocks can appear in different spaces. Intuitively
motives behave therefore like fundamental particles. A brief
physics oriented discussion of motives and some of their
applications can be found in \cite{rs08}, and an extensive
mathematical treatment is contained in \cite{motives94}.

 Over the past hundred years it has become clear through much
 mathematical experimentation that motives $M$ often are automorphic
 in the sense that their L-functions $L(M,s)$, viewed as complex functions,
 are identical to L-functions associated to automorphic forms $L(\Phi,s)$
 $$
 L(M,s) ~=~ L(\Phi,s).
 $$
 The only rigorously established case is the geometric
 interpretation of modular forms of weight two with respect to
 congruence groups of level $N$ in terms of motives associated to
 elliptic curves of conductor $N$. First steps in this direction
 were taken by Klein and Hurwitz in the late 19th century, but a
 general result was proven only a century later by Wiles and Taylor
  for stable elliptic curves \cite{w95etal}, and in complete
  generality in ref. \cite{bcdt01}. For higher dimensional motives
  the automorphic
 framework is more involved, and no general results are known.
 However, a number of conjectures about the expected structure of
 automorphic
 forms and representations have been formulated within the context
 of the Langlands program. This allows us to compare the
 conjectured motivic structure of Siegel modular forms with the
 compactification geometry of the CHL$_N$ model.

\subsection{Genus $n$ Siegel modular motives}

 The fact that the partition function of black holes in CHL$_N$ models
 is determined completely by genus two Siegel modular forms $\Phi^N$ implies that the
 problem of finding a geometric origin of the entropy of CHL$_N$ black holes
 translates into the question whether it is possible to reverse engineer for each
 of the forms $\Phi^N$ one or several modular motives $M_N$ such that
 the resulting motivic modular forms lead to the Siegel forms in a
 canonical way.  The general philosophy of the Langlands program
 suggests that this question makes sense in the context of any physical
 automorphic form, but in general this is an unsolved problem.
 The web of conjectures formulated by Langlands and others makes it
 possible however to make some general observations about the geometric
 interpretation of Siegel modular forms without specifying the degree of
 supersymmetry, or the type of compactification associated to it.

 While the theory of automorphic forms is not developed far enough to
 allow universal statements even at the conjectural level, for
 Siegel modular forms of arbitrary genus $n$ the expected motives
 can be characterized in a way that is precise enough to evaluate their
 physical relevance. Two quantities that give numerical information about
 the ambient variety are the weight $\rmwt(M)$ and the rank $\rmrk(M)$
 of the motive $M$.
 The former determines the cohomology group $H^i(X)$ in which the
 cohomological realization $H(M)$ of the motive $M$ lives. For
 Siegel forms the natural generalization of classical modular motives is the
 spinor motive $M_\Phi$,
 which is characterized by its particular form of the L-function
  in terms of the Satake parameters.
  \index{Satake parameters}
 For forms of weight $w$ and genus $n$ the weight  $\rmwt(M_\Phi)$
 of the induced spinor motive
 $M_\Phi$ can be read off the expected functional
 equation to be given by
  \beq
  \rmwt(M_\Phi) ~=~ nw - \frac{n}{2}(n+1),
  \eeq
  while the rank of the Siegel modular motive
   \index{Siegel modular motive}
  can be read off the spinor L-function as given by
  \beq
  \rmrk(M_\Phi) ~=~ 2^n.
  \eeq
  Furthermore, the Hodge decomposition of the cohomological
  realization $H(M_\Phi)$ of $M_\Phi$ is given for genus two forms
  by
  \beq
  H(M_\Phi) ~=~ H^{2w-3,0} \oplus H^{w-1,w-2} \oplus H^{w-2,w-1}
   \oplus H^{0,2w-3}.
  \eeq

 These results show that even though for genus two Siegel forms the rank
 of the motive is realistic within the framework of Calabi-Yau
 threefold compactifications, the weight of genus two Siegel modular
  motives leads to varieties of dimension
  $\rmdim_\mathC X = 2w-3$. For several of the CHL$_N$ models the dimension is
 therefore too high.

 We have seen above that the CHL$_N$ Siegel forms are determined
 via a canonical construction in terms of Maa\ss-Skoruppa roots
 $f^N$ that are classical modular forms. It is therefore natural to
 ask whether these forms lead to modular motives that can be
 accommodated by string compactifications.

\subsection{Motives of classical modular forms}

The fact that the Siegel forms $\Phi^N$ are determined completely
in terms of classical modular forms $f^N$ via the Maa\ss-Skoruppa
lift shows that the essential information pertaining to the
different models for varying $N$ is completely contained in the
forms $f^N$. Asking for a geometric origin of the $\Phi^N$
therefore mean to find a geometric interpretation of $f^N$.

Classical modular forms $f \in S_w(\G_0(N),\e_N)$ can be viewed as
genus one forms, leading to motives $M_f$ of weight
 \beq
 \rmwt(M_f) ~=~ w-1
 \eeq
 and rank $\rmrk(M_f)=2$, with Hodge structure given by
 \beq
  H(M_f) ~=~ H^{w-1,0} \oplus H^{0,w-1}.
 \eeq
 In the case of the CHL$_N$ models compactified on $X_N$  the question
 becomes how one can determine motives $M_N$ as a combination
 of $K3,E,N$. We therefore can ask the more concrete question whether the
 modular forms involved in the black hole entropy are determined by
 motives of the type
 \beq
 M_N ~=~ M(K3,E,N)
 \eeq
 possibly including the precise form of the action of the group
 $\mathZ_N$.

The Maa\ss-Skoruppa roots $f^{N}$, describing the purely
electrically charged states of the CHL$_N$ models, do not appear
to follow any obvious pattern that would suggest a geometric
origin. The range of modular weights is between three and twelve,
which makes a direct spacetime interpretation via Calabi-Yau
varieties again impossible for similar reasons that Siegel modular
motives cannot be physical for the CHL$_N$ black holes. While the
Hodge structure $H^{w-1,0}\oplus H^{0,w-1}$ does exist in any
Calabi-Yau variety, the dimension of the variety supporting the
motive is too high for the CHL$_N$ models because the
cohomological realization $H(M_\Om)$ of the motive is defined by a
Galois orbit in the intermediate cohomology
 $H(M_\Om) \subset H^n(X_n)$, where $n=\rmdim_\mathC X_n$.
 If the motive is pure the weight $w$ of the corresponding modular
 form $f_M$ (if it exists) is given by the complex dimension as
 $w(f_M) ~=~ n+1$.
For the CHL$_N$ models this implies that a direct Calabi-Yau
interpretation of the Maa\ss-Skoruppa root would have to involve
manifolds of complex dimensions ranging from two to eleven. As a
result most of the dimensions that appear in the CHL$_N$ models
are too high to be derived from a compactification space within
string theory, M-theory, or F-theory.

What turns the black hole entropy of CHL$_N$ into a probe that is
manageable is the fact that the CHL$_N$ Siegel modular forms can
be built from modular forms of weight two, thereby extending the
Maa\ss-Skoruppa lift one step further, as we will show in the next
section. Our constructions work by adding to the lift diagram
(\ref{ms-lift}) one further reduction $f_2^\tN \lra f^N$.

\vskip .3truein

 \section{The motivic origin of CH$_N$ black hole entropy}

The fact that neither the Siegel forms $\Phi^N$ nor their
Maa\ss-Skoruppa roots $f^N$ lead to physical motives raises the
question whether the Maa\ss-Skoruppa lift can be pushed further,
i.e. whether the forms $f^N$ can in turn be constructed in some
way from even simpler building blocks, and if so, whether those
building blocks admit a geometric interpretation. Given the
structure of the extra dimensions in the CHL$_N$ models it would
be natural to expect that the ingredients in such a construction
might involve modular forms of weight two, associated to elliptic
curves, or modular forms of weight three, associated to K3
surfaces, or both. It is this problem that we address in this
section.

It turns out that the sequence of classical black hole forms given
by the Maa\ss-Skoruppa roots splits into two distinct and disjoint
classes of forms, hence no completely universal reduction should
be expected. The fact that these classes form disjoint sets
guarantees that the elliptic reductions we describe are unique (up
to isogeny). The property that distinguishes certain of the
 forms $f^N$ is concerned with their symmetry structure. While the
majority of the black hole forms $f^N \in S_{w+2}(\G_0(N), \e_N)$
have no particular symmetry, the forms at levels $N=4,7,8$ admit a
particularly simple structure because they are of complex
multiplication (CM) type. We will answer the question raised above
about the geometric origin of the CHL$_N$ entropy by showing that
all CHL$_N$ Maa\ss-Skoruppa roots $f^{N}$ can be constructed in
terms of classical modular forms of weight two $f_2^\tN(q) \in
S_2(\G_0(\tN))$ via two different reductions, one for the non-CM
type forms, the other for the CM type forms. These modular forms
are supported by the unique motives of elliptic curves $E_\tN$ of
conductor $\tN$ that are determined up to isogeny. We describe
these two reductions in the following subsections.

\subsection{Non-CM elliptic reduction of CHL$_N$ Maa\ss-Skoruppa roots}

Our first reduction applies to those  CHL$_N$ models for which the
black hole root $f^N$ has no CM, i.e. $N=1,2,3,5,6$.  The key
observation here is that for these models we write the higher
weight classical black hole modular forms $f^{N}(q)$ defining the
Maa\ss-Skoruppa roots in terms of classical modular forms of
weight two as follows. For each order $N$ as given, there exists
an integer $\wtN$ and a cusp form of level $\tN$ and weight two,
$f_2^\tN(q) \in S_2(\G_0(\tN))$, such that the classical black
hole form $f^{N} \in S_{w+2}(\G_0(N),\e_N)$ can be written as
 \beq
 f^{N}(q) ~=~ f_2^\tN(q^{1/m})^m
 \lleq{non-cm-red}
 where
 \beq
 m= \frac{1}{2} \left\lceil\frac{24}{N+1}\right\rceil.
 \eeq
 The relation between the order $N$ of the CHL$_N$ group $\mathZ_N$
 and the level $\wtN$ of the weight two form is given in Table 2.
 \begin{center}
 \begin{tabular}{l| c c c c c}
 Order $N$        &1 &2 &3 &5 &6 \tabroom \\
 \hline

 Level $\wtN$ &36 &32 &27  &20 &24 \tabroom \\

 \hline
 \end{tabular}
 \end{center}

 \centerline{{\bf Table 2.}~{\it The levels $\wtN$ in terms of
 the orders $N$ of the CHL$_N$ models.}}

The construction (\ref{non-cm-red}) of the Maa\ss-Skoruppa root
from building blocks given by classical modular forms of weight
two then leads to a geometric interpretation in terms of elliptic
curves by applying the elliptic modularity theorem proven by
Wiles, Taylor and others \cite{w95etal}. This theorem proves the
conjecture Taniyama-Shimura-Weil conjecture, according to which
every elliptic curve over the rational numbers is modular in the
sense that its L-function agrees with the L-function of a weight
two form. Weil in particular made this somewhat vague conjecture
more precise by his important experimental observation that for an
elliptic curve $E_N$ of conductor $N$ the associated modular form
of weight 2 is associated to a Hecke congruence group $\G_0(N)$ at
level $N$. Given our weight two modular forms $f_2^\tN(q) \in
S_2(\G_0(\tN))$ derived from the black hole forms $f^N(q)$ we can
therefore construct elliptic curves $E_\tN$ of conductor $\tN$
such that their associated modular forms $f_2(E_\tN,q)$ are given
by the elliptic roots
 \beq
 f_2(E_\tN,q) ~=~ f_2^\tN(q).
 \eeq
  We will call the forms $f_2^\tN(q)$ the elliptic roots of the
  Siegel forms $\Phi^N$.
   The proof of this relation can be given explicitly via a
 case by case analysis for $N=1,2,3,5,6$, without the
 abstract machinery that enters Wiles' proof of
 the Taniyama-Shimura-Weil conjecture for stable elliptic curves,
 and the more general proof for all elliptic curves
 by Breuil, Conrad, Diamond and Taylor \cite{w95etal}.

 The reduction (\ref{non-cm-red}) turns out to lead to elliptic curves $E_\tN$
 which admit complex multiplication for $N=1,2,3$, and to curves with no
 CM for $N=5,6$.

 This leaves the CHL$_N$ models with $N=4,7,8$. For these cases the reduction
 (\ref{non-cm-red}) cannot be applied because the Maa\ss-Skoruppa
 roots $f^{N}$ have odd weight. For this class of forms it is
 necessary to introduce a different type of reduction, to which we
 turn in the next subsubsection.

\subsection{CM type elliptic reduction}

 As mentioned above, the Maa\ss-Skoruppa roots $f^{N}$ for the CHL$_N$
 models fall into two different classes of different symmetry types.
 While for $N=1,2,3,5,6$ these forms do not exhibit any particular symmetries,
 for $N=4,7,8$ these forms admit complex multiplication. As a consequence,
 these forms are sparse in the sense that their coefficients $a_p$ vanish
  for half the primes. It is this property which we will use to complete
  our elliptic reduction for the remaining CHL$_N$ models.

 There are several ways to think about complex multiplication
 forms $f(\tau)\in S_w(\G_0(N),\e)$. The point of view that explains
 the vanishing behavior of its Fourier coefficients in the most
 direct way is encoded in the definition originally given by
 Ribet \cite{r77}. The key here is that associated to each CM form is an
 imaginary quadratic field $K_D=\mathQ(\sqrt{-D})$, with $D$
 square free, such that the coefficients $a_p$, for $p$ prime, of its Fourier series
  $f(q)=\sum_n a_nq^n$, vanish for
 precisely those rational primes $p$ that do not split in the ring
 of integers $\cO_K$ of the field $K_D$. The splitting behavior of
 the rational primes within $\cO_{K_D}$ is controlled by the Legendre
 symbol $\chi_D$: if $\chi_D(p)=1$ then the prime $p$ factors in
 $\cO_{K_D}$. A CM form $f\in S_w(\G_0(N),\e)$ therefore can be defined
 through its expansion by the condition that there exists a field
 $K_D$ such that
 \beq
 \chi_D(p) a_p ~=~ a_p.
 \eeq

 To make the elliptic reduction of the black hole forms $f^{N}$ for the
 CHL$_N$ models
 with $N=4,7,8$ more transparent it is useful to shift
 perspective,
 and to consider the $L-$functions associated to the modular forms $f^N$,
 defined by the Mellin transform. Given the Fourier expansion
 $f(q) = \sum_n a_nq^n$
 of any cusp form $f\in S_w(\G_0(N),\e)$,
 the Mellin transform associates to $f$ the $L-$series
 \beq
 L(f,s) ~=~ \sum_n \frac{a_n(f)}{n^s}.
 \eeq
 The fact that the modular forms $f^{N}(q)$ for the CHL$_N$ models
 with $N=4,7,8$
  have complex multiplication means that their $L-$functions can be identified
  with the $L-$series of algebraic Hecke characters $\Psi_N$ associated to
  extensions $K_N$
  of the rational field $\mathQ$. In the present discussion
 the relevant fields are imaginary quadratic extensions $K_N=\mathQ(\sqrt{-D_N})$,
 where $D_N$
 is a square free integer.

 Algebraic Hecke characters associated to imaginary quadratic fields are
 defined by a congruence ideal $\mfrak \subset \cO_K$.
 If $\cI_\mfrak$ denotes the
 fractional ideals prime to $\mfrak$, algebraic Hecke characters are maps
 \beq
 \Psi:~ \cI_\mfrak ~\lra ~ \mathC^\times
 \eeq
 that can be normalized to be given on the principal ideals as
 \beq
 \Psi((z))~=~ z^w, ~~~{\rm with}~~z\equiv 1(\rmmod^\times \mfrak),
 \eeq
 where $\rmmod^\times$ denotes multiplicative congruence. The
 integer $w$ denotes the weight of $\Psi$.

 Given an algebraic Hecke character $\Psi$ of conductor $\cfrak_{\Psi}$,
 associated to an
 imaginary field $K$ of discriminant $-D$, define integers $a_n$ by
 summing over all
 integral ideals $\afrak$ of $K$ coprime to $\cfrak_\Psi$ as
 \beq
 a_n =  \sum_{\stackrel{(\afrak,\cfrak_{\Psi})=1}{\rmN\afrak=n}}
   \Psi(\afrak),
 \eeq
 where $\rmN\afrak$ denotes the norm of the ideal $\afrak$.
 The $L-$series of the character $\Psi$ is defined as
  \beq
   L(\Psi,s)  ~=~ \sum_{n=1}^{\infty} \frac{a_n(\Psi)}{n^s}.
  \eeq

 It is a theorem of Hecke that the $q-$series $f(\Psi,q) = \sum_n
 a_n(\Psi)q^n$, with $q=e^{2\pi i \tau}$, associated to $L(\Psi,s)$ via
 the inverse Mellin transform defines a modular form of weight $(w+1)$,
 level
 \beq
 N_{\Psi} ~=~ D\rmN \cfrak_{\Psi},
 \eeq
 and its Nebentypus character given by
 \beq
 \e_{\Psi}(m) ~=~ \frac{1}{m^w} \chi_D(m) \Psi((m)).
 \eeq
 Here $\chi_D(m)$ is the Legendre symbol defined above.
 i.e. $f(\Psi,q) \in S_{w+1}(\G_0(N_\Psi), \e_\Psi)$.

For the CHL$_N$ models with groups of orders $N=4,7,8$, the
imaginary quadratic fields $K_N= \mathQ(\sqrt{-D_N})$ are given by
the Gauss field $\mathQ(\sqrt{-1})$ and $\mathQ(\sqrt{-7})$,
$\mathQ(\sqrt{-2})$, respectively. These fields are all of class
number one, hence all the ideals are principal.

The CM nature of the modular forms $f^N \in S_{w+2}(\G_0(N),\e_N)$
implies that the essential information of these forms is encoded
in modular forms of weight 2 at levels $\tN$ that depend on the
order $N$ of the quotient groups $\mathZ_N$. This can be seen as
follows. First, there exist algebraic Hecke characters $\Psi_\wtN$
such that the $L-$series of the black hole forms are given by
powers $\Psi^{w+1}_\tN$ of the characters $\Psi_\tN$
  \beq
  L(f^N, s) ~=~ L(\Psi^{w+1}_\tN,s),
  \lleq{cm-red}
 for $(N,\tN) =(4,32),(7,49),(8,256)$.
 The elliptic origin of the classical black hole forms $f^N$
 now follows from the fact that the $L-$function $L(\Psi_\wtN,s)$ of
 the Hecke
 character $\Psi_\wtN$ is the Mellin transforms of a weight two modular form
 $f_2^\tN \in S_2(\G_0(\tN)$. It can furthermore be shown that the
 Mellin transform
 of these forms $f_2^\tN$ agree with the $L-$series of elliptic curves
 $E_\wtN$ with conductor $\wtN$
 which admit complex multiplication  by $K_D$
  \beq
  L(E_\tN,s) ~=~ L(\Psi_\tN,s) ~=~ L(f_2^\tN,s).
  \lleq{cm-ell-curves}
 In this way we obtain a second systematic construction that leads to an elliptic
 curve interpretation for the remaining CHL$_N$ models, thereby
 completing our elliptic reduction of the Siegel forms
 $\Phi^N$ for all CHL$_N$ models.

 The Hecke $L-$series interpretation (\ref{cm-red}) of the black hole forms
 $f^N$ for $N=4,7,8$,
 combined with (\ref{cm-ell-curves}), gives
 the most systematic formulation of the link between the high weight forms
 $f^N$ and the weight two forms
 $f_2^\tN$.  For the examples of the present paper it is possible to express
 the coefficients
 of the black hole forms $f^{N}$ in a more direct, but less transparent, way
 in terms of the coefficients of the classical weight two forms
 $f_2^\wtN$. Expanding the forms as
 \beq
 f^{N}(q) ~=~ \sum_n b^{w+2}_n q^n,~~~~~~~ f_2^\wtN(q) ~=~ \sum_n a_n q^n
 \eeq
 one can derive for the coefficients at primes $p$ the relations
 \bea
  b_p^3 &=& a_p^2 - 2p \nn \\
  b_p^5 &=& a_p^4 - 4pa_p^2 + 2p^2.
 \llea{symmetric-powers}
The relevant elliptic forms of weight two for $N=4,7,8$ are given
in Table 3. For $N=7,8$ we use the symbol $\cong$ to indicate that
we are listing only the coefficients $a_p$ for primes $p$. These
are the important coefficients because all other Fourier
coefficients can be obtained via the Hecke relations
 \bea
  a_{p^n} &=& a_pa_{p^{n-1}} + p^wa_{p^{n-2}} \nn \\
  a_{mn} &=& a_m a_n, ~~~{\rm for}~m\neq n,
 \eea
 because the
forms of weight two are Hecke eigenforms.  Combining these results
with those of Table 2 gives the conductors $\tN$ of the elliptic
curve $E_\tN$ associated to each of the CHL$_N$ model associated
to a quotient manifold with respect to $\mathZ_N$.

 \begin{tabular}{c| c l}
 $(w+2, N,\wtN)$   &$K$  &$f_2^\wtN \in S_2(\G_0(\wtN))$ \tabroom \\
\hline
 $(5, 4,32)$  &$\mathQ(\sqrt{-1})$ &$f_2^{32}(q) ~=~ \eta(q^4)^2\eta(q^8)^2$ \tabroom \\
 $(3, 7,49)$  &$\mathQ(\sqrt{-7})$ &$f_2^{49}(q) ~\cong ~
                           q + q^2 + 4q^{11} + 8q^{23} + 2q^{29}
                              - 6q^{37} -12q^{43}  + \cdots$  \tabroom \\
            % - 10q^{53}
 $(3, 8,256)$  &$\mathQ(\sqrt{-2})$ &$f_2^{256}(q) ~\cong~
                        q + 2q^3 + 6q^{11} - 6q^{17} + 2q^{19} + 6q^{41}
                                   - 10q^{43}  + \cdots$ \tabroom \\
             %+ 6q^{59}
 \hline
\end{tabular}

\vskip .1truein

 \centerline{{\bf Table 3.} ~{\it Elliptic forms
associated to the CHL$_N$ models for $N=4,7,8$.}}

 By considering the expansion of the
congruence group black hole forms $f^{N}$ for $N=4,7,8$ it can be
checked that the coefficients their $L-$functions are given in
terms of the relations (\ref{symmetric-powers}). We will call this
construction of the high weight Maa\ss-Skoruppa roots $f^N$ in
terms of classical modular forms of weight two the CM-type
elliptic reduction.

 \subsection{Elliptic weight two roots of the CHL$_N$ Maa\ss-Skoruppa
 roots}

As a  result of our two lift constructions (\ref{non-cm-red}) and
(\ref{cm-red}), we have established that the geometric origin of
the black hole forms $\Phi^N \in S_w(\G_0^{(2)}(N))$ is mediated
by the interpretation of the classical forms $f^N \in
S_{w+2}(\G_0(N),\e_N)$ in terms of weight two forms $f_2^\tN\in
S_2(\G_0(\tN))$.  The elliptic curves $E_\tN$ whose motives
 support the modular forms $f_2^\tN$ are determined up to isogeny, i.e.
  maps that are surjective and have a finite kernel. The results
of this geometric interpretation are summarized in Table 4, which
also includes an indication whether the forms that appear are of
complex multiplication type or not.
 \begin{center}
 \begin{small}
 \begin{tabular}{c |  c l | c l c}

 $N$ of     &   &BH Form       &   &Motivic form       &Level $\wtN$  \\
 CHL$_N$    &   &$f^N(q) \in S_{w+2}(\G_0(N))$
                               &   &$f_2^\wtN(q)$      &of $E_{\wtN}$ \\

 \hline
  1         &   &$\eta(\tau)^{24}$
                               &CM    & $\eta(q^6)^4 \in S_2(\G_0(36))$
                                                       &16  \tabroom  \\

 \hline

  2         &    &$\eta(\tau)^8\eta(2\tau)^8$
     &CM     &$\eta(q^4)^2 \eta(q^8)^2 \in S_2(\G_0(32))$  &32  \tabroom  \\
    %     &          &CM: $\eta(q^4)^2 \eta(q^8)^2\otimes \chi_2 \in S_2(\G_0(64))$ & 64 \\

 \hline

  3         &   &$\eta(\tau)^6\eta(3\tau)^6$
                       &CM   &$\eta(q^3)^2 \eta(q^9)^2$
                                     &27 \tabroom \\

  \hline

  4         &CM  & $\eta(\tau)^4\eta(2\tau)^2\eta(4\tau)^4$
                           &   &$\rmSym^4(f_2^{32})$ with $f_2^{32} \in S_2(\G_0(32))$
                            & 32  \tabroom \\

  \hline

  5         &   &$\eta(\tau)^4\eta(5\tau)^4$
                           &  &$\eta(q^2)^2 \eta(q^{10})^2 \in S_2(\G_0(20))$
                           &20  \tabroom   \\

  \hline

  6         &  &$(\eta(\tau)\eta(2\tau)\eta(3\tau)\eta(6\tau))^2$
       &  &$\eta(2\tau)\eta(4\tau)\eta(6\tau)\eta(12\tau) \in S_2(\G_0(24))$
                          & 24 \tabroom  \\

  \hline

  7         &CM  &$\eta(\tau)^3\eta(7\tau)^3$
         &    &$\rmSym^2f_2^{49}$ with $f_2^{49} \in S_2(\G_0(49))$
                                    & 49 \tabroom \\

 \hline

  8          &CM  &$\eta(\tau)^2\eta(2\tau)\eta(4\tau)\eta(8\tau)^2$
         &   &$\rmSym^2f_2^{256}$ with $f_2^{256} \in S_2(\G_0(256))$
                                    & 256 \tabroom \\

 \hline
 \end{tabular}
 \end{small}
 \end{center}

\centerline{{\bf Table 4.}~{\it Motives associated to the electric
modular forms of CHL$_N$ models.}}

We see from the reductions compiled in Table 4 that the elliptic
lifts $f_2^\tN \lra f^N$ of the CHL$_N$ modular forms lead to
classical modular forms of weight two that admit CM for $N=1,2,3$.
Combining this with the three Maa\ss-Skoruppa roots $f^N$ at
$N=4,7,8$ leaves the two forms at $N=5,6$ that do not have CM.
This indicates that CM is not a fundamental property as far as the
geometric structure of these models is concerned. We show in the
Appendix that it is possible to construct these two non-CM forms
in terms of non-geometric forms of weight 1 that do admit complex
multiplication.

 \vskip .3truein

\section{Conclusions}

In this paper we have described a program to view black holes as
probes of the geometry of extra dimensions $-$ given a
hypothetical black hole in the laboratory, we can ask what the
data extracted from this black hole might tell us about the
details of the small scale structure of spacetime. We have shown
that this strategy can be made concrete in the context of
automorphic black holes by combining it with the idea that
automorphic forms are supported by motives. While the precise
framework of such automorphic motives is not known at present,
certain concrete features are expected to be present for motives
that support such forms.

Assuming that the conjectured properties of automorphic motives we
see that in the context of Siegel automorphic black hole entropy
the motives induced by the Siegel forms are not physical for
dimensional reasons. Nor are the motives induced by the
Maa\ss-Skoruppa roots which count $\frac{1}{2}$-BPS states for the
same reason. The key to the geometry is the realization that the
Maa\ss-Skoruppa roots that appear in the class of CHL$_N$ models
are in fact of a special type such that they can be constructed
from modular cusp form of weight two. We have shown that
independent of the complex multiplication properties of the
Maa\ss-Skoruppa root it is possible to induce the black hole roots
$f^N$ from forms $f_2 \in S_2(\G_0(\tN))$ with $\tN(N)$, leading
to a lift diagram that extends the usual Maa\ss-Skoruppa lift
 $$
  f_2^\tN ~\lra ~f^N ~\stackrel{\rmSk}{\lra} \vphi^N
       ~ \stackrel{\rmMS}{\lra} ~\Phi^N,
 $$
 where we have reinstated the weight $(w+2)$ given by
 (\ref{weight-of-ms-roots}).
 These forms of weight two in turn are supported by elliptic curves.

The fact that the Siegel forms $\Phi^N$ are induced by elliptic
motives shows that the Siegel count of the CHL$_N$ black hole
entropy only contains a limited amount of information about the
geometry of spacetime.  It would be of interest to refine the
black hole formulae so as to encode more detailed motivic
information that allows to reconstruct spacetime more precisely.
 A possible strategy in this direction would be to consider more
 general black holes
 \cite{dgn07} than have been considered so far for the class of
 CHL$_N$ models.

\vskip .2truein

{\large {\bf Acknowledgement.}} \hfill \break
 It is a pleasure to thank Atish Dabholkar,
  Rajesh Gopakumar, and Monika Lynker for discussions.
  The authors are grateful for support of this research by the National
  Science Foundation
  under grant No. PHY 0969875. RS thanks the Schr\"odinger Institute
  for hospitality.

\vskip .3truein

\section{Appendix: CM properties of black hole entropy}

The list of forms in Table 4 shows that three out of the 8 CHL$_N$
Maa\ss-Skoruppa roots that describe the small black holes in these
models are of CM type, namely those at $N=4,7,8$. The elliptic
forms of weight two obtained for the remaining classical forms via
the non-CM type reduction (\ref{non-cm-red}) are of CM type for
$N=1,2,3$, leading to elliptic curves with complex multiplication
symmetry. This leaves two forms and their associated elliptic
curves without CM, and raises the question there these non-CM type
modular forms cannot be built in some other way from forms that do
admit CM, if perhaps in a non-geometric way. It turns out that the
answer is affirmative.

The remaining two forms at $N=5,6$ can be constructed in
  terms of CM forms of weight one as
  \bea
  f_{1,80}(\tau) &=& \eta(4\tau)\eta(20\tau) \in S_1(\G_1(80)) \nn \\
  f_{1,128}(\tau) &=& \eta(8\tau)\eta(16\tau) \in S_1(\G_1(128)).
  \eea
 With these forms we can write
  \bea
   f_2^{20}(q) &=& f_1^{80}(q^{1/2})^2 \nn \\
   f_2^{24}(q) &=& f_1^{128}(q^{1/4})f_{1,128}(q^{3/4}).
  \eea
  This shows that if one considers reductions to non-geometric modular forms of weight 1
   all the black hole modular forms can be constructed in terms of CM modular forms.

 \vskip .3truein

%\parskip=0.05truein
%\baselineskip=21pt

% \hfill First draft, South Bend, December 11, 2010

%\printindex


\begin{thebibliography}{9}
 \bibitem{chl95} S. Chaudhuri, G. Hockney and J.D. Lykken, {\it
 Maximally supersymmetric string theories in $D<10$}, Phys. Rev.
 Lett. {\bf 75} (1995) 2264, arXiv: hep-th/9505054
   \bibitem{dvv96} R. Dijkgraaf, E.P. Verlinde and H.L. Verlinde,
 {\it Counting dyons in $N=4$ string theory}, Nucl. Phys. {\bf B484} (1997)
 543 $-$ 561, arXiv: hep-th/9607026
  \bibitem{js05} D.P. Jatkar and A. Sen, {\it Dyon spectrum in CHL models},
           JHEP {\bf 04} (2006) 018, arXiv: hep-th/0510147
  \bibitem{gk09} S. Govindarajan and K.G. Krishna, {\it BKM Lie superalgebras
        from dyon spectra in $\mathZ_N$ CHL orbifolds for composite $N$},
         JHEP {\bf 05} (2010) 014, arXiv: 0907.1410 [hep-th]
 \bibitem{ssy05} D. Shih, A. Strominger and X. Yin, {\it Recounting dyons
      in $N=4$ string theory}, JHEP {\bf 10} (2005) 087, arXiv: hep-th/0505094
   \bibitem{rs08}  R. Schimmrigk, {\it Emergent spacetime from
   modular motives}, Commun. Math. Phys. {\bf 303} (2011) 1 $-$ 30,
     arXiv: 0812.4450 [hep-th]
  \bibitem{as05a} A. Sen, {\it Black holes, elementary strings
  and holomorphic anomaly}, JHEP {\bf 07} (2005) 063, arXiv:
  hep-th/0502126
 \bibitem{m79} H. Maa\ss, {\it \"Uber eine Spezialschar von Modulformen
  zweiten Grades} I$-$III, Invent. Math. {\bf 52} (1979) 95 $-$
  104, {\bf 53} (1979) 249 $-$ 253, 255 $-$ 265
   \bibitem{nps92} N. Skoruppa, {\it Computations of Siegel modular forms of genus two},
             Math. Comp. {\bf 58} (1992) 381 $-$ 398
 \bibitem{mrv93} M. Manickam, B. Ramakrishnan and T.C. Vasudevan,
      {\it On Saito-Kurokawa descent for forms of degree two},
      Manuscripta Math. {\bf 81} (1993) 161 $-$ 182
  \bibitem{cg08} F. Cl\'ery and V.A. Gritsenko, {\it The Siegel modular forms of
         genus 2 with the simplest divisor}, arXiv: 0812.3962 [math.NT]
  \bibitem{motives94} U. Jannsen, S. Kleiman and J.-P. Serre, {\it Motives}, Proc.  of Symposia in
   Pure Mathematics, Vol. 55.1 and 55.2, Amer. Math. Soc.,  1994
  \bibitem{w95etal} A. Wiles, {\it Modular elliptic curves and Fermat's last theorem},
            Ann. Math {\bf 141} (1995) 443 $-$ 551; \\
          R. Taylor and A. Wiles, {\it Ring theoretic properties of certain Hecke
            algebras}, Ann. Math. {\bf 141} (1995) 553 $-$ 572
 \bibitem{bcdt01} C. Breuil, B. Conrad, F. Diamond and R. Taylor,
             {\it On the modularity of elliptic curves over $\mathQ$: wild 3-adic exercises},
              Amer. J. Math (2001)
 \bibitem{r77} K.A. Ribet, {\it Galois representations attached to eigenforms with nebentypus},
        in {\sc Modular functions of one variable V}, eds. J.-P. Serre and D. Zagier, LNM 601,
        Springer, 1977
  \bibitem{cdwkm04} G.L Cardoso, B. de Wit, J. Kappeli and T.
 Mohaupt, {\it Asymptotic degeneracy of dyonic $\cN=4$ string states
 and black hole entropy}, JHEP {\bf 12} (2004) 075, arXiv:
 hep-th/0412287
  \bibitem{dgn07} A. Dabholkar, D. Gaiotto and S. Nampuri, {\it Comments on the spectrum
          of CHL dyons}, JHEP {\bf 01} (2008) 023, arXiv: hep-th/0702150
\end{thebibliography}
\end{document}